\documentclass[aip,reprint]{revtex4-1}
\usepackage{physics}
\usepackage{graphicx}
\usepackage{dcolumn}
\usepackage[hidelinks]{hyperref}
\usepackage{bm}
\usepackage[dvipsnames]{xcolor}

\draft 

\begin{document}

\preprint{}

\title{Acoustic forces near elastic substrate}

\author{V. Kleshchenko}

\affiliation{Qingdao Innovation and Development Center, Harbin Engineering University, Qingdao 266000 Shandong, China}
\affiliation{School of Physics and Engineering, ITMO University, 197101, Saint-Petersburg, Russia}%
 \author{K. Albitskaya}
\affiliation{School of Physics and Engineering, ITMO University, 197101, Saint-Petersburg, Russia}%
 \affiliation{Moscow Institute of Physics and Technology, Dolgoprudny, 141700, Russia}
 \affiliation{Russian Quantum Center, Skolkovo, 143025 Moscow, Russia}
\author{M. Petrov}
 \email{m.petrov@metalab.ifmo.ru}
\affiliation{School of Physics and Engineering, ITMO University, 197101, Saint-Petersburg, Russia}%
\affiliation{Qingdao Innovation and Development Center, Harbin Engineering University, Qingdao 266000 Shandong, China}

\date{\today}

\begin{abstract}
In this work, we study the acoustic forces acting on particles due to sound scattering at the interface with an elastic substrate. Utilizing the Green's function formalism, we predict that excitation of leaking Rayleigh wave results in strong modification of the acoustic pressure  force acting on a monopole scatterer and changes  the equilibrium position of particles above the substrate surface. We also showed that the presence of a substrate changes the configuration of the acoustical binding of two particles due to multiple rescattering of acoustic wave from the interface. The reported results propose the method of acoustic manipulation via surface waves excitation and    demonstrate the effect from elastic media in acoustical trapping of microobjects. 
\end{abstract}


\maketitle


The mechanical forces generated by acoustical fields  are widely utilized  for trapping and binding of   macro- and microparticles in various fields, including biology and biophysics\cite{Ozcelik2018, Meng2019, Dholakia2020, drinkwater2020perspective}. In particular, acoustic tweezers are used for cell trapping\cite{yin2023acoustic, baresch2016observation, li2021acoustic}, sorting \cite{fan2022recent, ahmed2018vertical}, and acoustic levitation\cite{andrade2020acoustic, pazos2022particle}. The nature of the acoustic forces lies in the transfer of momentum between the sound field and the matter during scattering process. 

Strong acoustical forces can act on particles ensembles when the scattered fields from each particle interfere with the external field. Multiple scattering leads to appearance of various acoustomechanical effects  such as, for instance,  acoustic binding effect\cite{bjerknes_fields_1906, Doinikov1995, Silva2014, mohapatra2018experimental, Sepehrirahnama2015, Lopes2016, hoque2020interparticle, Habibi2017, Tang2021, Zhang2016, pavlic2022interparticle, Eslami2023, Baasch2017, StClair2023, wang2017sound, Silva2019} when particles form  stable spatial configurations in the potential of the effective scattered fields. The acoustical binding  was firstly studied for the gas bubbles \cite{bjerknes_fields_1906, Doinikov1995}. Later, the studies dedicated to the investigation of the acoustic forces in ensembles of particles, including sets of small  \cite{Silva2014, mohapatra2018experimental, hoque2020interparticle} or comparable to the wavelength scatterers \cite{Sepehrirahnama2015, Lopes2016, Habibi2017, wang2017sound, Tang2021, Silva2019}, and, in particular, distantly interacting spheres \cite{Zhang2016} were carried out. In this context, the viscous effects\cite{sepehrirahnama2016effects, Eslami2023, pavlic2022interparticle} and  dynamical properties \cite{Baasch2017, StClair2023} were also  explored.


\begin{figure}[t]
    \includegraphics[width=0.4\textwidth]{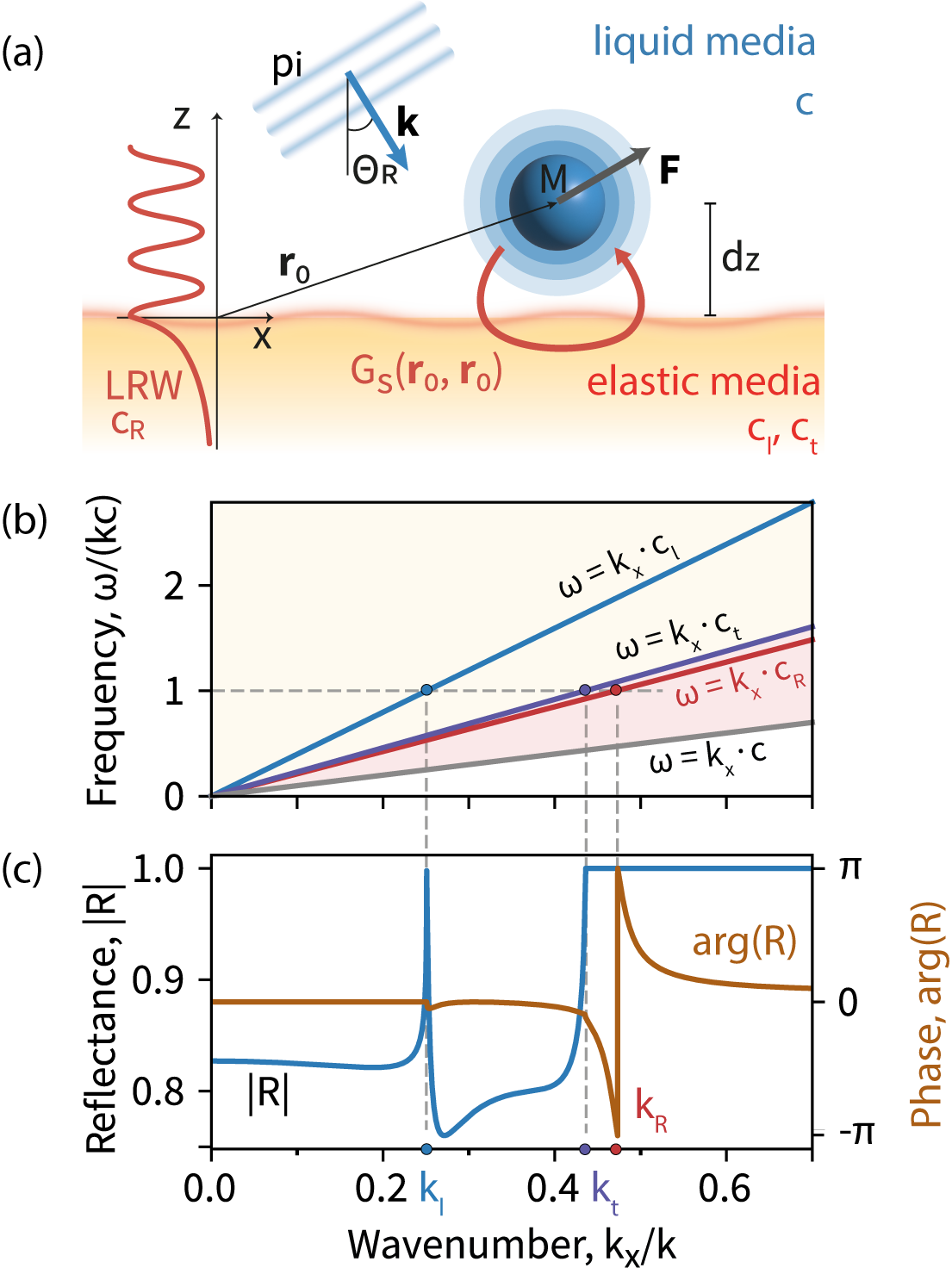}
    \caption{\label{fig:1} (a) Scattering of a plane wave incident at a Rayleigh angle on a single particle and schematic field profile of leaky Rayleigh wave. (b) Dispersion of longitudinal and transversal waves in substrate (blue and purple lines correspondingly), leaky Rayleigh wave (red line) and longitudinal waves in the upper medium (grey line). (c) Reflection coefficient of plane wave at the water (liquid media)-quartz (elastic media) interface depending on parallel component of wavevector.}
\end{figure}

Mostly, acoustic binding and multiple scattering effects were discussed in the scenario of  free space scattering in gaseous or liquid environment. However, in many practical applications the spatial boundaries with solid interfaces should be taken into account, especially, when it comes to biophysical applications in lab-on-chip devices. Since recently acoustic trapping and sensing near solid interface became one of the most prominent directions of research due to surface acoustic waves (SAW) which can be excited at the solid-liquid interface to manipulate the objects\cite{collins2015two, ahmed2018vertical, nguyen2018patterning, mandal2022surface, destgeer2013continuous}. These waves are commonly generated by piezoelectric interdigital transducers and propagate in the form of leaky Rayleigh waves (LRW) when part of the energy is radiated  to the liquid half-space.  The surface waves propagate in two dimensions and decay more slowly than bulk waves, which makes them more effective for manipulating particles over long distances.  For example, by using the interference of multiple propagating SAWs, it becomes possible to achieve spatial control over particles located near the solid surface. \cite{collins2015two}. 

In this work, we study the acoustical force acting on individual microparticles and their pairs in the vicinity of a substrate that supports LRW (see Fig.~\ref{fig:1} (a)). We build a theoretical approach basing on  the Green's function formalism \cite{Toftul2019, williams1999fourier} and extend it to the case of acoustical binding in the manner similar to optical binding effect\cite{Kostina2019}. We analyze the contributions of SAW into the acoustical force acting on particles near a substrate  and show that these waves can significantly alter the acoustomechanical interactions.



When an scatterer is placed at a relatively small distance to the substrate,  multiple scattering effects between the particles and the surface occur. Various numerical \cite{Wang2014, Baasch2020, Khler2022} and analytical \cite{Bostrm1980, Gaunaurd1994, Shenderov2002, Miri2011, Wang2012, Qiao2017, Mitri2018, Zang2019, Maksimov2020, Simon2023} studies addressed this problem considering  spherical particles\cite{Gaunaurd1994, Shenderov2002, Miri2011, Zang2019, Baasch2020, Simon2023}, cylindrical\cite{Wang2012, Qiao2017, Mitri2018, Khler2022} objects and bubbles\cite{Wang2014, Doinikov2015, Maksimov2020}. Most of the theoretical models are based on the images methods when the particle reflection is introduced in order to describe interaction with the substrate. However, this approximation does not completely account for  waves excitation in some fluid or elastic substrates  and multiple reflections (see Supplementary materials Sec. I). Below we build a consistent model of acoustic forces acting on a  monopole scatterer near an elastic half-space.

We start by considering a compressible particle  placed in a medium near an elastic substrate in the external pressure field $p_i(\vb r)$. The medium is characterized by mass density $\rho$, compressibility $\beta$ and the speed of sound $c=1/\sqrt{\rho\beta}$. The substrate density is $\rho_s$ and sound velocities of longitudinal and transverse waves are $c_l$ and $c_t$ respectively (Fig.~\ref{fig:1}a).

In the most general case time-averaged acoustic radiation force acting on an object can be obtained by integrating of the acoustic stress tensor\cite{Westervelt1957} over the surface surrounding this object $\partial \Omega$:
\begin{eqnarray}
    \mathbf{F} = -\int_{\partial \Omega} \left< 
    \frac12 \left( \beta |p|^2 - \rho |\mathbf{v}|^2 \right) \mathbf{\hat{I}} + \rho \mathbf{v}^* \mathbf{v}
    \right> \cdot \mathbf{n} dS,
\end{eqnarray}
where $p$, $\mathbf{v}$ are pressure and velocity local fields.
However, for the particle placed at the coordinate $\mathbf{r}_0$, pressure field in the monopole approximation is given by\cite{williams1999fourier} $p(\mathbf{r}) = p_i(\mathbf{r}) + i \rho \omega G(\mathbf{r} , \mathbf{r}_0) M$ and expression for the force reduces to the simpler form \cite{Toftul2019}: 
\begin{eqnarray}
    \mathbf{F} = -\frac{1}{2\omega}\text{Im} (M^*\nabla p).
    \label{eq:fmon}
\end{eqnarray}
Here, $M$ is the monopole moment of a scatterer, and $G = G_0 + G_s$ is the sum of the free-space Green function and it's reflected from a substrate part (see more details in Supplementary materials Sec. II):
\begin{eqnarray}
    \label{eq:gs}
    G_s(\mathbf{r}, \mathbf{r}_0)
    = \frac{i}{4\pi} \int_0^{\infty} \!\!\!\!\!\! R(k_x) e^{ik_z(z + d_z)} J_0(k_x d_x) \frac{k_x}{k_z} \dd k_x.
\end{eqnarray}
$G_s$ is represented using the plane wave expansion as the integral over the parallel component of wavevector $k_x$, where $k_z = \sqrt{k^2 - k^2_x}$, $k=\omega/c$ is the wavenumber in medium, $R(k_x)$ is the reflection coefficient of plane wave from elastic halfspace, $d_x=|x-x_0|$, $d_z=z_0$. 

The monopole moment $M$ can be defined through the monopole polarizability of the particle $\alpha$:
\begin{eqnarray}
    \label{eq:mmom}
    M = -i\omega \beta \alpha p(\mathbf{r}_0),
\end{eqnarray}
which describes it's scattering properties and, in the case of spherical particle in a free space, can be expressed in terms of the monopole component of T-matrix \cite{williams1999fourier, sapozhnikov2013radiation, Lopes2016, Toftul2019} $\alpha_0 = - i{4\pi} t_0/{k^3}$. If the particle is placed near a substrate, the following expression for the monopole polarizability can be  obtained after the renormalizaiton procedure \cite{Kostina2019}:
\begin{eqnarray}
    \label{eq:efpol}
    \alpha^s = \left[1 - k^2 G_s(\mathbf{r}_0, \mathbf{r}_0) \alpha_0 \right]^{-1} \alpha_0,
\end{eqnarray}

The factor gives a correction for the free space polarizability of the particle and   accounts for multiple scattering between particle and the substrate. This can be seen more evidently by writing down all the contributions of the scattered waves series \cite{DeVries1998}. The same factor can be obtained from the solution of the equation on self-consisted scattered field as discussed in more  details  in Supplementary materials  Sec. III A.

In this approach, the parameters of the substrate are accounted for in the interface reflection amplitude $R(\theta)$ which defines the amplitude of the plane  wave reflected from the solid substrates incident at the angle    $\theta$. For the case of sound rigid substrate $R=1$, however in the case of elastic substrate it depends on the angle and frequency of the incidence wave  or, in other words, (as in Eq.~\ref{eq:gs}) on the parallel component of the wavevector $k_x$ \cite{maurice_ewing_elastic_1957, royer1999elastic}: 
\begin{eqnarray}
    R(k_x) = \frac{R_1 - iR_2}{R_1 + iR_2},
    \label{eq:reflectance}
\end{eqnarray}
where $R_1 = (2 k^2_x - k^2_t)^2 k_z - 4 k^2_x (k_x^2-k_l^2)^{\frac12} (k_x^2-k_t^2)^{\frac12} k_z$, $R_2 = k^4_{t}(k_x^2-k_l^2)^{\frac12} \rho/\rho_s$ and $k_l=\omega/c_l$, $k_t=\omega/c_t$.

The absolute value and phase of the reflection amplitude are plotted in Fig.~\ref{fig:1}(c). Also, the dispersion lines for four types of waves are also shown in Fig.~\ref{fig:1}b, with the slope defined by the corresponding sound velocities normalized on $c$.  The reflection coefficient has singularities at wavenumbers correspondent to excitation of the waves in elastic media. The first singularity is reached at the value $k_x=k_l$ correspondent to longitudinal waves propagating along the surface, and the absolute value  of the  reflection coefficient reaches unity. Next, for wavenumbers larger than $k_x=k_t$, both longitudinal and transverse waves become evanescent in the elastic media, and no power is transmitted to the solid ($|R|=1$). At the wavenumber $k_x=k_R$, which is typically slightly higher than $k_t$, the rapid $2\pi$-phase shift is observed. This singularity is a  pole of reflection amplitude  (Eq.\ref{eq:reflectance}) in the complex $k_x$-plane~\cite{bertoni1973unified} and corresponds  to the excitation of LRW. This is a surface acoustic wave that is evanescent in the substrate and propagates along the solid-fluid interface as $\exp(ik_Rx)$. The propagation speed of the wave is $c_R=\omega/\text{Re}(k_R)$. It radiates into the fluid half-space at Rayleigh angle $\theta_R$ (see Fig.~\ref{fig:1} (a)), and the attenuation is determined by the non-zero imaginary part of the wavevector $\text{Im}(k_R)$. There is another mechanism of dissipation originating from frictional losses because of the transverse motion of the surface, however this effect is typically smaller, especially at higher frequencies \cite{gedge2012acoustofluidics}.

\begin{figure}[t]
    \includegraphics[width=0.45\textwidth]{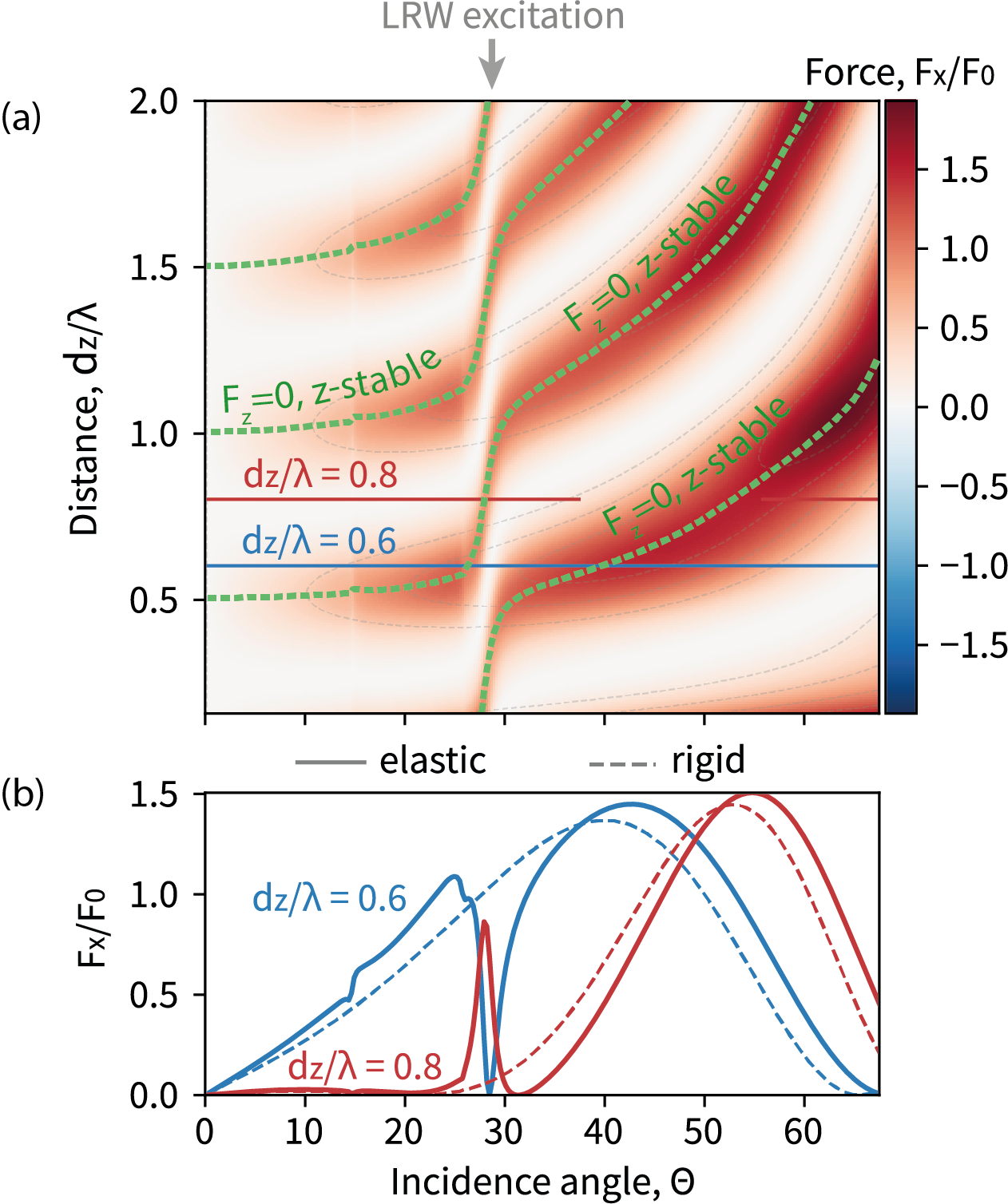}
    \caption{\label{fig:2} (a) $x$-component of the acoustic force depending on the angle of incidence and height above the elastic surface. Dashed lines denote the case of sound rigid substrate. The green dashed lines show the conditions for stable position along the $z$-coordinate. (a) The cross-section of the map at $d_z/\lambda=0.6(0.8)$ for the case of elastic substrate and rigid surface.}
\end{figure}

Excitation of the bulk and surface waves in the elastic substrate contributes to the resulting acoustic force. To explore this effect, we considered polystyrene particle $\rho_p=1045$ kg/$\text{m}^3$, $c_p=1020$ m/s with radius $a=100$ $\mu$m placed in water ($\rho=1000$ kg/$\text{m}^3$, $c=1480$ m/s) near a quartz substrate with $\rho_s=2650$ kg/$\text{m}^3$, $c_l=5900$ m/s, $c_t=3400$ m/s. The incident plane wave had frequency $\approx2.4$ MHz ($ka=1$). 
In Fig.~\ref{fig:2} (a), the normalized  $x$-component of acoustic force, acting on particle depending on the angle of incidence is presented for different distances from the substrate. The force is normalized over $F_0 = \frac12 \beta p_0^2 \pi a^2$, which is pressure force by plane acoustic wave on a disk with the geometrical cross section of the particle.  One can see that the force has minima and maxima correspondent to the nodes and anti-nodes of the formed standing wave. Thin grey dashed lines correspond to the force level lines for the sound rigid substrate. One can see that the elastic and rigid substrate behave quite similar but in the narrow region close to the to bulk and surface wave excitation.  In particular, close to LRW excitation at  $\theta_R\approx28^{\circ}$ the abrupt change of the $x$-component of the force down to almost zero values is observed at the distance $d_z/\lambda=0.6$ (see Fig.~\ref{fig:2} (b)). This happens due to resonant excitation of LRW and additional 2$\pi$ phase shift of the reflected acoustic wave. At other distance from the substrate,i.e. $d_z/\lambda=0.8$, this results in resonant enhancement of the $x$-force, red line in Fig.~\ref{fig:2} (b). The peculiarities near the bulk waves excitation are also present however they are much weaker. One should also note  that LRW excitation  drastically changes the stable position along $z$-direction (green dashed line in ~\ref{fig:2} (a)) where accumulation of additional $2\pi$ phase at the resonant is clearly seen.  Thus, excitation of LRW can resonantly enhance or suppress the acoustic pressure force.

\begin{figure*}
    \includegraphics[width=\textwidth]{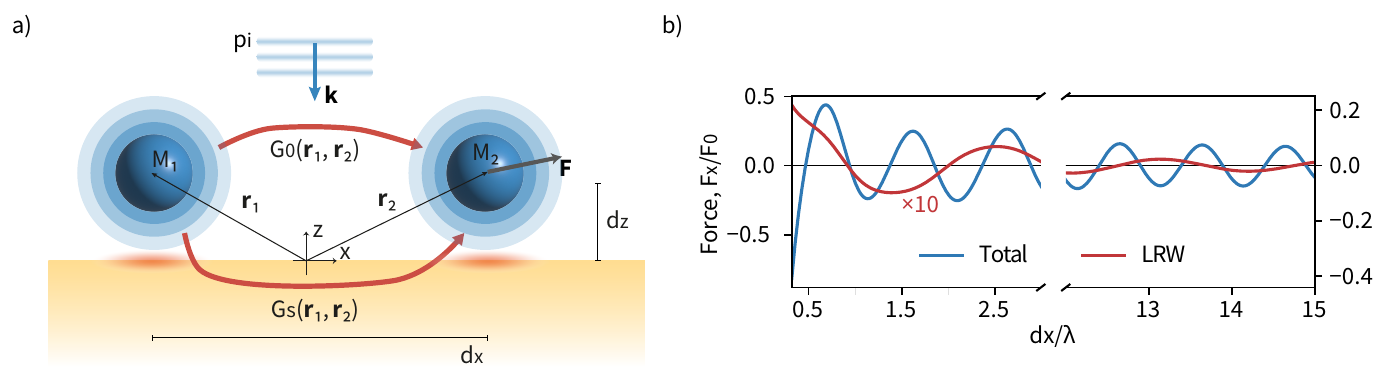}
    \caption{\label{fig:3}  (a) acoustic binding of two interacting monopole particles near the substrate in the field of normally incident plane wave. (b) binding force depending on distance between the particles. Blue line represents the total $x$-component of acoustic force acting on right particle. Red line stands for the contribution to the force from the Leaky Rayleigh wave excited by scatterers. The results are shown for particles located at height of $d_z/\lambda=0.5$ and frequency(size) parameter $ka=1$.}
\end{figure*}

Next, we analyze how elastic substrate can alter the acoustical binding. For that  we examined the system of two particles placed in the coordinates $\mathbf{r}_1, \mathbf{r}_2$ as shown in Fig.~\ref{fig:3} (a). In this case, local pressure field consists of external field $p_i$ and total scattered fields from each of the particles:
\begin{eqnarray}
    p = p_i + i\rho \omega \, G(\mathbf{r}, \mathbf{r}_1)M_1 + i\rho \omega \, G(\mathbf{r}, \mathbf{r}_2)M_2,
\end{eqnarray}
where $M_1$ and $M_2$ are the monopole moments of the first and the second scatterer. Here, it is also necessary to take into account the effects of rescattering between particles  in addition to the multiple reflections between the particles and the elastic halfspace. Analogously to the case of one particle interacting with the substrate, this can be accounted via the the effective polarizabilities as discussed in more details in Supplementary materials Sec. III B:
\begin{eqnarray}
    \tilde\alpha_i^s = \left[1 - k^4 G(\mathbf{r}_i, \mathbf{r}_j) \alpha_j^s G(\mathbf{r}_j, \mathbf{r}_i) \alpha_i^s \right]^{-1} \alpha_i^s.
\end{eqnarray}
In this expression, which is very similar to  Eq.~\eqref{eq:efpol}, $\alpha_i^s$ is polarizability of $i$-th single particle in presence of the substrate introduced earlier. 
In other words, we find the exact solution of the scattering problem on monopole particles arbitrary placed  near the substrate in external field. The current solution accounts explicitly for the multiple scattering effects between the particles as well as multiple reflections from the elastic substrate. Additionally, this general expression can be used for interacting particles in free space by simply substituting the free space polarizabilities instead of $\alpha_i^s$.

For the the binding effect, we consider a plane wave incident normally to the surface while particles are placed at the same height above the substrate. In this configuration due to symmetry of the problem, there is no  $x$-component of the net force acting on two particles and they can form  stable binded configuration.  For the fixed position of monopoles along $z$-axis $d_z/\lambda\approx0.5$ the dependence of $x$-components of the force on distance between particles $d_x/\lambda$ is presented at Fig.~\ref{fig:3} (b). This is an exclusively binding forces, which arises due to the rescattering effects between the particles. Zeros of the $x$-force components correspond to the alternating positions of stable and unstable equilibriums along $x$-axis. The amplitude of the force decreases as the distance between spheres increases. This behaviour is determined by Eq.~\ref{eq:fmon}, which shows the distance dependence of the monopole moments (polarizabilities) of the particles and the gradient of the scattered pressure field, given by the gradient of the Green's function.

\begin{figure}
    \includegraphics[width=0.5\textwidth]{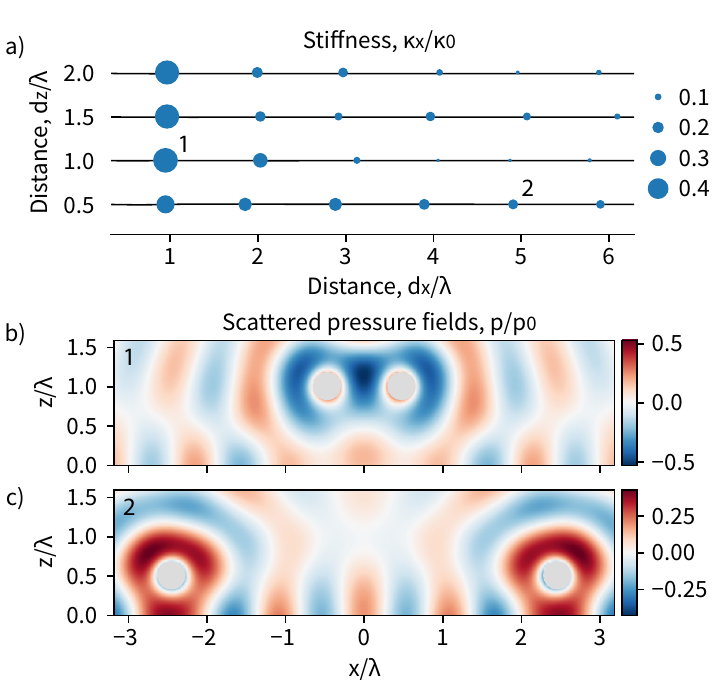}
    \caption{\label{fig:4} (a) stable positions of particles pair determined by the height above the surface $d_z/\lambda$ and the distance between the scatterers $d_x/\lambda$. Circles sizes represent normalized binding stiffness in equilibrium positions. (b), (c) normalized scattered pressure fields ... in stable configurations 1 and 2 correspondingly, and labeled in Fig.~\ref{fig:4} (a).}
\end{figure}
In opposite to the free space model, scattered fields of the particles not only act on the other particles through the bulk waves in the fluid, but also interact with a substrate, producing additional coupling channel between the objects. Beside the interaction by ordinary reflected waves, binding through the surface waves becomes possible in case of elastic substrate. However, our calculations show that this contribution is rather weak and binding through the bulk waves in liquid prevails. The contribution to the force components from the fields that are coupled to the SAWs is also shown in Fig.~\ref{fig:3}b. It can be found as the residue at the point $k_x=k_R$ of the integrand originating from gradient of Green's function (Eq.~\ref{eq:gs}) that is used for evaluation corresponding force part (Eq.~\ref{eq:fmon}):
\begin{eqnarray}
    F_{R,x} = \frac{\rho_0}{8} \text{Re} \big[ M_1^*
        I(k_R) \text{res}R|_{k_R} M_2 \big].
\end{eqnarray}
Derivation of this formula is given in Supplementary materials IV. It's worth noting that the SAW-based contribution to the force component is small.

Finally, we determine the equilibrium positions of  two particles along both $x$- and $z$-axes. We fix the pair of particles at the same height (see Fig.~\ref{fig:3}a) and compute the $x$- and $z$-components of the force in full range of configurations. We first found the stable configuration along $z$-coordinate which are predominantly determined by the nodes of the standing wave formed by the incident external field and its reflection from the substrate. The stable planes in $z$-direction are denoted with horizontal lines in Fig.~\ref{fig:4} (a). The stable configurations of a pair of scatterers along $x$-axis are shown with blue circles. The radius of the circle in Fig.~\ref{fig:4} (a) denotes the binding stiffness along $x$-axis  $\kappa_x$ defined as $F_x = -\kappa_x \Delta x$ and normalized over $\kappa_0=F_0/R$. It can be seen, that the binding stiffness is larger for small distances between particles due to the stronger scattered field. However,   the binding stiffness depends on both $x$ and $z$ coordinates in non-trivial manner which can be explained by the complex pattern of the acoustic fields close to the surface formed by scattering from the surface. This is also supported by the fact that the stable positions are not equidistantly spaced like contrary to homogeneous environment.  Two particular stable geometries of binded particles for $d_z/\lambda\approx1$ and $d_x/\lambda\approx1$ (configuration 1) and $d_z/\lambda\approx0.5$ and $d_x/\lambda\approx5$ (configuration 2) are shown in Fig.~\ref{fig:4} (b) demonstrating the complex structure of the scattered fields. It is also noticeable that the acoustic field is enhanced in the gap between the particles and the solid surface once particles are located close to it.

While the effects proposed and discussed in this paper have not been observed experimentally yet, there are setups and geometries which can offer potential experimental verification. As such, the control and separation of particles with SAW have been experimentally investigated in several different setups \cite{ahmed2018vertical, destgeer2013continuous}. 
Acoustic interparticle forces were recently measured by recording dynamics of particles levitating within a plane \cite{garcia2014experimental,mohapatra2018experimental}, moving along channels \cite{hoque2020interparticle}, or being trapped close to another fixed particle \cite{saeidi2019acoustic}. It should be noted that these measurements were made far from the spatial boundaries and mostly for small particles. Recently, an experimental study which is very close to the proposed geometry was carried out\cite{simon2023measurement}. In this work, the radiation force acting on a polypropylene sphere in water near a boundary was measured by forming a pendulum with the sphere near a wall and tracking its displacement in the field of acoustic source \cite{nikolaeva2016measuring}. While there was no influence of excited surface acoustic waves observed in that particular geometry, it could be observed for an incline plane wave incidence. In addition, the measured effects were examined for large particles, $ka>10$, where the monopole approximation is not applicable. Nevertheless, we believe that the reported technique can help in the experimental verification of the effects proposed in this paper. We also stress that the limitations of the proposed model such as neglecting gravity and buoyant forces, viscosity, and higher order multiples contribution will not affect the predicted effects as discussed in more detail in Supplementary Materials Sec. V.

Finally, the proposed approach can be extended to other localized acoustic waves such as waves at the acoustic metamaterial interfaces. Due to the negative effective parameters of metamaterial substrate, one can expect excitation of the evanescent surface waves~\cite{bliokh2019transverse, park2011amplification, ambati2007surface}, which could lead to a significant enhancement of the acoustomechanical effects at small distances above the substrate. Thus, the reported results demonstrate the role of the elastic substrate and surface acoustic wave in modification of acoustic forces, and, as we believe, may reconsider the role of localized acoustic waves for mechanical manipulation of microobjects.

Refer to the supplementary materials for further details on the derivation of analytical expressions for the Green's function, effective polarizabilities, and the acoustic force contribution from the leaky Rayleigh wave, as well as additional discussion on the advantages and limitations of the proposed theoretical model.

We thank Konstantin Ladutenko, Ivan Toftul, and Natalia Kostina for discussion and valuable advises. The work was supported by the Academic Leadership Program Priority 2030. The analytical simulations were supported bu the Russian Science Foundation (project number 22-42-04420).  The numerical simulations were supported by Ministry of Science and Higher Education of the Russian Federation (Agreement No. 075-15-2021-1349).


%
%

%



\bibliography{my-bib}

\providecommand{\noopsort}[1]{}\providecommand{\singleletter}[1]{#1}%
\begin{thebibliography}{63}%
\makeatletter
\providecommand \@ifxundefined [1]{%
 \@ifx{#1\undefined}
}%
\providecommand \@ifnum [1]{%
 \ifnum #1\expandafter \@firstoftwo
 \else \expandafter \@secondoftwo
 \fi
}%
\providecommand \@ifx [1]{%
 \ifx #1\expandafter \@firstoftwo
 \else \expandafter \@secondoftwo
 \fi
}%
\providecommand \natexlab [1]{#1}%
\providecommand \enquote  [1]{``#1''}%
\providecommand \bibnamefont  [1]{#1}%
\providecommand \bibfnamefont [1]{#1}%
\providecommand \citenamefont [1]{#1}%
\providecommand \href@noop [0]{\@secondoftwo}%
\providecommand \href [0]{\begingroup \@sanitize@url \@href}%
\providecommand \@href[1]{\@@startlink{#1}\@@href}%
\providecommand \@@href[1]{\endgroup#1\@@endlink}%
\providecommand \@sanitize@url [0]{\catcode `\\12\catcode `\$12\catcode
  `\&12\catcode `\#12\catcode `\^12\catcode `\_12\catcode `\%12\relax}%
\providecommand \@@startlink[1]{}%
\providecommand \@@endlink[0]{}%
\providecommand \url  [0]{\begingroup\@sanitize@url \@url }%
\providecommand \@url [1]{\endgroup\@href {#1}{\urlprefix }}%
\providecommand \urlprefix  [0]{URL }%
\providecommand \Eprint [0]{\href }%
\providecommand \doibase [0]{http://dx.doi.org/}%
\providecommand \selectlanguage [0]{\@gobble}%
\providecommand \bibinfo  [0]{\@secondoftwo}%
\providecommand \bibfield  [0]{\@secondoftwo}%
\providecommand \translation [1]{[#1]}%
\providecommand \BibitemOpen [0]{}%
\providecommand \bibitemStop [0]{}%
\providecommand \bibitemNoStop [0]{.\EOS\space}%
\providecommand \EOS [0]{\spacefactor3000\relax}%
\providecommand \BibitemShut  [1]{\csname bibitem#1\endcsname}%
\let\auto@bib@innerbib\@empty
\bibitem [{\citenamefont {Ozcelik}\ \emph {et~al.}(2018)\citenamefont
  {Ozcelik}, \citenamefont {Rufo}, \citenamefont {Guo}, \citenamefont {Gu},
  \citenamefont {Li}, \citenamefont {Lata},\ and\ \citenamefont
  {Huang}}]{Ozcelik2018}%
  \BibitemOpen
  \bibfield  {author} {\bibinfo {author} {\bibfnamefont {A.}~\bibnamefont
  {Ozcelik}}, \bibinfo {author} {\bibfnamefont {J.}~\bibnamefont {Rufo}},
  \bibinfo {author} {\bibfnamefont {F.}~\bibnamefont {Guo}}, \bibinfo {author}
  {\bibfnamefont {Y.}~\bibnamefont {Gu}}, \bibinfo {author} {\bibfnamefont
  {P.}~\bibnamefont {Li}}, \bibinfo {author} {\bibfnamefont {J.}~\bibnamefont
  {Lata}}, \ and\ \bibinfo {author} {\bibfnamefont {T.~J.}\ \bibnamefont
  {Huang}},\ }\bibfield  {title} {\enquote {\bibinfo {title} {Acoustic tweezers
  for the life sciences},}\ }\href {\doibase 10.1038/s41592-018-0222-9}
  {\bibfield  {journal} {\bibinfo  {journal} {Nature Methods}\ }\textbf
  {\bibinfo {volume} {15}},\ \bibinfo {pages} {1021–1028} (\bibinfo {year}
  {2018})}\BibitemShut {NoStop}%
\bibitem [{\citenamefont {Meng}\ \emph {et~al.}(2019)\citenamefont {Meng},
  \citenamefont {Cai}, \citenamefont {Li}, \citenamefont {Zhou}, \citenamefont
  {Niu},\ and\ \citenamefont {Zheng}}]{Meng2019}%
  \BibitemOpen
  \bibfield  {author} {\bibinfo {author} {\bibfnamefont {L.}~\bibnamefont
  {Meng}}, \bibinfo {author} {\bibfnamefont {F.}~\bibnamefont {Cai}}, \bibinfo
  {author} {\bibfnamefont {F.}~\bibnamefont {Li}}, \bibinfo {author}
  {\bibfnamefont {W.}~\bibnamefont {Zhou}}, \bibinfo {author} {\bibfnamefont
  {L.}~\bibnamefont {Niu}}, \ and\ \bibinfo {author} {\bibfnamefont
  {H.}~\bibnamefont {Zheng}},\ }\bibfield  {title} {\enquote {\bibinfo {title}
  {Acoustic tweezers},}\ }\href {\doibase 10.1088/1361-6463/ab16b5} {\bibfield
  {journal} {\bibinfo  {journal} {Journal of Physics D: Applied Physics}\
  }\textbf {\bibinfo {volume} {52}},\ \bibinfo {pages} {273001} (\bibinfo
  {year} {2019})}\BibitemShut {NoStop}%
\bibitem [{\citenamefont {Dholakia}, \citenamefont {Drinkwater},\ and\
  \citenamefont {Ritsch-Marte}(2020)}]{Dholakia2020}%
  \BibitemOpen
  \bibfield  {author} {\bibinfo {author} {\bibfnamefont {K.}~\bibnamefont
  {Dholakia}}, \bibinfo {author} {\bibfnamefont {B.~W.}\ \bibnamefont
  {Drinkwater}}, \ and\ \bibinfo {author} {\bibfnamefont {M.}~\bibnamefont
  {Ritsch-Marte}},\ }\bibfield  {title} {\enquote {\bibinfo {title} {Comparing
  acoustic and optical forces for biomedical research},}\ }\href {\doibase
  10.1038/s42254-020-0215-3} {\bibfield  {journal} {\bibinfo  {journal} {Nature
  Reviews Physics}\ }\textbf {\bibinfo {volume} {2}},\ \bibinfo {pages}
  {480–491} (\bibinfo {year} {2020})}\BibitemShut {NoStop}%
\bibitem [{\citenamefont {Drinkwater}(2020)}]{drinkwater2020perspective}%
  \BibitemOpen
  \bibfield  {author} {\bibinfo {author} {\bibfnamefont {B.~W.}\ \bibnamefont
  {Drinkwater}},\ }\bibfield  {title} {\enquote {\bibinfo {title} {A
  perspective on acoustical tweezers—devices, forces, and biomedical
  applications},}\ }\href@noop {} {\bibfield  {journal} {\bibinfo  {journal}
  {Applied Physics Letters}\ }\textbf {\bibinfo {volume} {117}} (\bibinfo
  {year} {2020})}\BibitemShut {NoStop}%
\bibitem [{\citenamefont {Yin}\ \emph {et~al.}(2023)\citenamefont {Yin},
  \citenamefont {Jiang}, \citenamefont {Mann}, \citenamefont {Tian},\ and\
  \citenamefont {Drinkwater}}]{yin2023acoustic}%
  \BibitemOpen
  \bibfield  {author} {\bibinfo {author} {\bibfnamefont {C.}~\bibnamefont
  {Yin}}, \bibinfo {author} {\bibfnamefont {X.}~\bibnamefont {Jiang}}, \bibinfo
  {author} {\bibfnamefont {S.}~\bibnamefont {Mann}}, \bibinfo {author}
  {\bibfnamefont {L.}~\bibnamefont {Tian}}, \ and\ \bibinfo {author}
  {\bibfnamefont {B.~W.}\ \bibnamefont {Drinkwater}},\ }\bibfield  {title}
  {\enquote {\bibinfo {title} {Acoustic trapping: An emerging tool for
  microfabrication technology},}\ }\href@noop {} {\bibfield  {journal}
  {\bibinfo  {journal} {Small}\ }\textbf {\bibinfo {volume} {19}},\ \bibinfo
  {pages} {2207917} (\bibinfo {year} {2023})}\BibitemShut {NoStop}%
\bibitem [{\citenamefont {Baresch}, \citenamefont {Thomas},\ and\ \citenamefont
  {Marchiano}(2016)}]{baresch2016observation}%
  \BibitemOpen
  \bibfield  {author} {\bibinfo {author} {\bibfnamefont {D.}~\bibnamefont
  {Baresch}}, \bibinfo {author} {\bibfnamefont {J.-L.}\ \bibnamefont {Thomas}},
  \ and\ \bibinfo {author} {\bibfnamefont {R.}~\bibnamefont {Marchiano}},\
  }\bibfield  {title} {\enquote {\bibinfo {title} {Observation of a single-beam
  gradient force acoustical trap for elastic particles: acoustical tweezers},}\
  }\href@noop {} {\bibfield  {journal} {\bibinfo  {journal} {Physical review
  letters}\ }\textbf {\bibinfo {volume} {116}},\ \bibinfo {pages} {024301}
  (\bibinfo {year} {2016})}\BibitemShut {NoStop}%
\bibitem [{\citenamefont {Li}\ \emph {et~al.}(2021)\citenamefont {Li},
  \citenamefont {Shen}, \citenamefont {Huang},\ and\ \citenamefont
  {Cummer}}]{li2021acoustic}%
  \BibitemOpen
  \bibfield  {author} {\bibinfo {author} {\bibfnamefont {J.}~\bibnamefont
  {Li}}, \bibinfo {author} {\bibfnamefont {C.}~\bibnamefont {Shen}}, \bibinfo
  {author} {\bibfnamefont {T.~J.}\ \bibnamefont {Huang}}, \ and\ \bibinfo
  {author} {\bibfnamefont {S.~A.}\ \bibnamefont {Cummer}},\ }\bibfield  {title}
  {\enquote {\bibinfo {title} {Acoustic tweezer with complex boundary-free
  trapping and transport channel controlled by shadow waveguides},}\
  }\href@noop {} {\bibfield  {journal} {\bibinfo  {journal} {Science advances}\
  }\textbf {\bibinfo {volume} {7}},\ \bibinfo {pages} {eabi5502} (\bibinfo
  {year} {2021})}\BibitemShut {NoStop}%
\bibitem [{\citenamefont {Fan}\ \emph {et~al.}(2022)\citenamefont {Fan},
  \citenamefont {Wang}, \citenamefont {Ren}, \citenamefont {Lin},\ and\
  \citenamefont {Wu}}]{fan2022recent}%
  \BibitemOpen
  \bibfield  {author} {\bibinfo {author} {\bibfnamefont {Y.}~\bibnamefont
  {Fan}}, \bibinfo {author} {\bibfnamefont {X.}~\bibnamefont {Wang}}, \bibinfo
  {author} {\bibfnamefont {J.}~\bibnamefont {Ren}}, \bibinfo {author}
  {\bibfnamefont {F.}~\bibnamefont {Lin}}, \ and\ \bibinfo {author}
  {\bibfnamefont {J.}~\bibnamefont {Wu}},\ }\bibfield  {title} {\enquote
  {\bibinfo {title} {Recent advances in acoustofluidic separation technology in
  biology},}\ }\href@noop {} {\bibfield  {journal} {\bibinfo  {journal}
  {Microsystems \& Nanoengineering}\ }\textbf {\bibinfo {volume} {8}},\
  \bibinfo {pages} {94} (\bibinfo {year} {2022})}\BibitemShut {NoStop}%
\bibitem [{\citenamefont {Ahmed}\ \emph {et~al.}(2018)\citenamefont {Ahmed},
  \citenamefont {Destgeer}, \citenamefont {Park}, \citenamefont {Jung},\ and\
  \citenamefont {Sung}}]{ahmed2018vertical}%
  \BibitemOpen
  \bibfield  {author} {\bibinfo {author} {\bibfnamefont {H.}~\bibnamefont
  {Ahmed}}, \bibinfo {author} {\bibfnamefont {G.}~\bibnamefont {Destgeer}},
  \bibinfo {author} {\bibfnamefont {J.}~\bibnamefont {Park}}, \bibinfo {author}
  {\bibfnamefont {J.~H.}\ \bibnamefont {Jung}}, \ and\ \bibinfo {author}
  {\bibfnamefont {H.~J.}\ \bibnamefont {Sung}},\ }\bibfield  {title} {\enquote
  {\bibinfo {title} {Vertical hydrodynamic focusing and continuous
  acoustofluidic separation of particles via upward migration},}\ }\href@noop
  {} {\bibfield  {journal} {\bibinfo  {journal} {Advanced science}\ }\textbf
  {\bibinfo {volume} {5}},\ \bibinfo {pages} {1700285} (\bibinfo {year}
  {2018})}\BibitemShut {NoStop}%
\bibitem [{\citenamefont {Andrade}, \citenamefont {Marzo},\ and\ \citenamefont
  {Adamowski}(2020)}]{andrade2020acoustic}%
  \BibitemOpen
  \bibfield  {author} {\bibinfo {author} {\bibfnamefont {M.~A.}\ \bibnamefont
  {Andrade}}, \bibinfo {author} {\bibfnamefont {A.}~\bibnamefont {Marzo}}, \
  and\ \bibinfo {author} {\bibfnamefont {J.~C.}\ \bibnamefont {Adamowski}},\
  }\bibfield  {title} {\enquote {\bibinfo {title} {Acoustic levitation in
  mid-air: Recent advances, challenges, and future perspectives},}\ }\href@noop
  {} {\bibfield  {journal} {\bibinfo  {journal} {Applied Physics Letters}\
  }\textbf {\bibinfo {volume} {116}} (\bibinfo {year} {2020})}\BibitemShut
  {NoStop}%
\bibitem [{\citenamefont {Pazos~Ospina}\ \emph {et~al.}(2022)\citenamefont
  {Pazos~Ospina}, \citenamefont {Contreras}, \citenamefont {Estrada-Morales},
  \citenamefont {Baresch}, \citenamefont {Ealo},\ and\ \citenamefont
  {Volke-Sep{\'u}lveda}}]{pazos2022particle}%
  \BibitemOpen
  \bibfield  {author} {\bibinfo {author} {\bibfnamefont {J.~F.}\ \bibnamefont
  {Pazos~Ospina}}, \bibinfo {author} {\bibfnamefont {V.}~\bibnamefont
  {Contreras}}, \bibinfo {author} {\bibfnamefont {J.}~\bibnamefont
  {Estrada-Morales}}, \bibinfo {author} {\bibfnamefont {D.}~\bibnamefont
  {Baresch}}, \bibinfo {author} {\bibfnamefont {J.~L.}\ \bibnamefont {Ealo}}, \
  and\ \bibinfo {author} {\bibfnamefont {K.}~\bibnamefont
  {Volke-Sep{\'u}lveda}},\ }\bibfield  {title} {\enquote {\bibinfo {title}
  {Particle-size effect in airborne standing-wave acoustic levitation: Trapping
  particles at pressure antinodes},}\ }\href@noop {} {\bibfield  {journal}
  {\bibinfo  {journal} {Physical Review Applied}\ }\textbf {\bibinfo {volume}
  {18}},\ \bibinfo {pages} {034026} (\bibinfo {year} {2022})}\BibitemShut
  {NoStop}%
\bibitem [{\citenamefont {Bjerknes}(1906)}]{bjerknes_fields_1906}%
  \BibitemOpen
  \bibfield  {author} {\bibinfo {author} {\bibfnamefont {V.~V.}\ \bibnamefont
  {Bjerknes}},\ }\href {http://archive.org/details/fieldsofforce00bjeruoft}
  {{\emph {\bibinfo {title} {Fields of force}}}}\
  (\bibinfo  {publisher} {New York, Col. Univ. Press},\ \bibinfo {year}
  {1906})\BibitemShut {NoStop}%
\bibitem [{\citenamefont {Doinikov}\ and\ \citenamefont
  {Zavtrak}(1995)}]{Doinikov1995}%
  \BibitemOpen
  \bibfield  {author} {\bibinfo {author} {\bibfnamefont {A.~A.}\ \bibnamefont
  {Doinikov}}\ and\ \bibinfo {author} {\bibfnamefont {S.~T.}\ \bibnamefont
  {Zavtrak}},\ }\bibfield  {title} {\enquote {\bibinfo {title} {On the mutual
  interaction of two gas bubbles in a sound field},}\ }\href {\doibase
  10.1063/1.868506} {\bibfield  {journal} {\bibinfo  {journal} {Physics of
  Fluids}\ }\textbf {\bibinfo {volume} {7}},\ \bibinfo {pages} {1923–1930}
  (\bibinfo {year} {1995})}\BibitemShut {NoStop}%
\bibitem [{\citenamefont {Silva}\ and\ \citenamefont
  {Bruus}(2014)}]{Silva2014}%
  \BibitemOpen
  \bibfield  {author} {\bibinfo {author} {\bibfnamefont {G.~T.}\ \bibnamefont
  {Silva}}\ and\ \bibinfo {author} {\bibfnamefont {H.}~\bibnamefont {Bruus}},\
  }\bibfield  {title} {\enquote {\bibinfo {title} {Acoustic interaction forces
  between small particles in an ideal fluid},}\ }\href {\doibase
  10.1103/physreve.90.063007} {\bibfield  {journal} {\bibinfo  {journal}
  {Physical Review E}\ }\textbf {\bibinfo {volume} {90}} (\bibinfo {year}
  {2014}),\ 10.1103/physreve.90.063007}\BibitemShut {NoStop}%
\bibitem [{\citenamefont {Mohapatra}, \citenamefont {Sepehrirahnama},\ and\
  \citenamefont {Lim}(2018)}]{mohapatra2018experimental}%
  \BibitemOpen
  \bibfield  {author} {\bibinfo {author} {\bibfnamefont {A.~R.}\ \bibnamefont
  {Mohapatra}}, \bibinfo {author} {\bibfnamefont {S.}~\bibnamefont
  {Sepehrirahnama}}, \ and\ \bibinfo {author} {\bibfnamefont {K.-M.}\
  \bibnamefont {Lim}},\ }\bibfield  {title} {\enquote {\bibinfo {title}
  {Experimental measurement of interparticle acoustic radiation force in the
  rayleigh limit},}\ }\href@noop {} {\bibfield  {journal} {\bibinfo  {journal}
  {Physical Review E}\ }\textbf {\bibinfo {volume} {97}},\ \bibinfo {pages}
  {053105} (\bibinfo {year} {2018})}\BibitemShut {NoStop}%
\bibitem [{\citenamefont {Sepehrirahnama}, \citenamefont {Lim},\ and\
  \citenamefont {Chau}(2015)}]{Sepehrirahnama2015}%
  \BibitemOpen
  \bibfield  {author} {\bibinfo {author} {\bibfnamefont {S.}~\bibnamefont
  {Sepehrirahnama}}, \bibinfo {author} {\bibfnamefont {K.-M.}\ \bibnamefont
  {Lim}}, \ and\ \bibinfo {author} {\bibfnamefont {F.~S.}\ \bibnamefont
  {Chau}},\ }\bibfield  {title} {\enquote {\bibinfo {title} {Numerical study of
  interparticle radiation force acting on rigid spheres in a standing wave},}\
  }\href {\doibase 10.1121/1.4916968} {\bibfield  {journal} {\bibinfo
  {journal} {The Journal of the Acoustical Society of America}\ }\textbf
  {\bibinfo {volume} {137}},\ \bibinfo {pages} {2614–2622} (\bibinfo {year}
  {2015})}\BibitemShut {NoStop}%
\bibitem [{\citenamefont {Lopes}, \citenamefont {Azarpeyvand},\ and\
  \citenamefont {Silva}(2016)}]{Lopes2016}%
  \BibitemOpen
  \bibfield  {author} {\bibinfo {author} {\bibfnamefont {J.~H.}\ \bibnamefont
  {Lopes}}, \bibinfo {author} {\bibfnamefont {M.}~\bibnamefont {Azarpeyvand}},
  \ and\ \bibinfo {author} {\bibfnamefont {G.~T.}\ \bibnamefont {Silva}},\
  }\bibfield  {title} {\enquote {\bibinfo {title} {Acoustic interaction forces
  and torques acting on suspended spheres in an ideal fluid},}\ }\href
  {\doibase 10.1109/tuffc.2015.2494693} {\bibfield  {journal} {\bibinfo
  {journal} {IEEE Transactions on Ultrasonics, Ferroelectrics, and Frequency
  Control}\ }\textbf {\bibinfo {volume} {63}},\ \bibinfo {pages} {186–197}
  (\bibinfo {year} {2016})}\BibitemShut {NoStop}%
\bibitem [{\citenamefont {Hoque}\ and\ \citenamefont
  {Sen}(2020)}]{hoque2020interparticle}%
  \BibitemOpen
  \bibfield  {author} {\bibinfo {author} {\bibfnamefont {S.}~\bibnamefont
  {Hoque}}\ and\ \bibinfo {author} {\bibfnamefont {A.}~\bibnamefont {Sen}},\
  }\bibfield  {title} {\enquote {\bibinfo {title} {Interparticle acoustic
  radiation force between a pair of spherical particles in a liquid exposed to
  a standing bulk acoustic wave},}\ }\href@noop {} {\bibfield  {journal}
  {\bibinfo  {journal} {Physics of Fluids}\ }\textbf {\bibinfo {volume} {32}}
  (\bibinfo {year} {2020})}\BibitemShut {NoStop}%
\bibitem [{\citenamefont {Habibi}, \citenamefont {Devendran},\ and\
  \citenamefont {Neild}(2017)}]{Habibi2017}%
  \BibitemOpen
  \bibfield  {author} {\bibinfo {author} {\bibfnamefont {R.}~\bibnamefont
  {Habibi}}, \bibinfo {author} {\bibfnamefont {C.}~\bibnamefont {Devendran}}, \
  and\ \bibinfo {author} {\bibfnamefont {A.}~\bibnamefont {Neild}},\ }\bibfield
   {title} {\enquote {\bibinfo {title} {Trapping and patterning of large
  particles and cells in a 1d ultrasonic standing wave},}\ }\href {\doibase
  10.1039/c7lc00640c} {\bibfield  {journal} {\bibinfo  {journal} {Lab on a
  Chip}\ }\textbf {\bibinfo {volume} {17}},\ \bibinfo {pages} {3279–3290}
  (\bibinfo {year} {2017})}\BibitemShut {NoStop}%
\bibitem [{\citenamefont {Tang}\ and\ \citenamefont {Huang}(2021)}]{Tang2021}%
  \BibitemOpen
  \bibfield  {author} {\bibinfo {author} {\bibfnamefont {T.}~\bibnamefont
  {Tang}}\ and\ \bibinfo {author} {\bibfnamefont {L.}~\bibnamefont {Huang}},\
  }\bibfield  {title} {\enquote {\bibinfo {title} {Mie particle assembly by a
  converging ultrasound field and acoustic interaction forces},}\ }\href
  {\doibase 10.1016/j.apacoust.2021.108123} {\bibfield  {journal} {\bibinfo
  {journal} {Applied Acoustics}\ }\textbf {\bibinfo {volume} {180}},\ \bibinfo
  {pages} {108123} (\bibinfo {year} {2021})}\BibitemShut {NoStop}%
\bibitem [{\citenamefont {Zhang}\ \emph {et~al.}(2016)\citenamefont {Zhang},
  \citenamefont {Qiu}, \citenamefont {Wang}, \citenamefont {Ke},\ and\
  \citenamefont {Liu}}]{Zhang2016}%
  \BibitemOpen
  \bibfield  {author} {\bibinfo {author} {\bibfnamefont {S.}~\bibnamefont
  {Zhang}}, \bibinfo {author} {\bibfnamefont {C.}~\bibnamefont {Qiu}}, \bibinfo
  {author} {\bibfnamefont {M.}~\bibnamefont {Wang}}, \bibinfo {author}
  {\bibfnamefont {M.}~\bibnamefont {Ke}}, \ and\ \bibinfo {author}
  {\bibfnamefont {Z.}~\bibnamefont {Liu}},\ }\bibfield  {title} {\enquote
  {\bibinfo {title} {Acoustically mediated long-range interaction among
  multiple spherical particles exposed to a plane standing wave},}\ }\href
  {\doibase 10.1088/1367-2630/18/11/113034} {\bibfield  {journal} {\bibinfo
  {journal} {New Journal of Physics}\ }\textbf {\bibinfo {volume} {18}},\
  \bibinfo {pages} {113034} (\bibinfo {year} {2016})}\BibitemShut {NoStop}%
\bibitem [{\citenamefont {Pavlic}, \citenamefont {Ermanni},\ and\ \citenamefont
  {Dual}(2022)}]{pavlic2022interparticle}%
  \BibitemOpen
  \bibfield  {author} {\bibinfo {author} {\bibfnamefont {A.}~\bibnamefont
  {Pavlic}}, \bibinfo {author} {\bibfnamefont {L.}~\bibnamefont {Ermanni}}, \
  and\ \bibinfo {author} {\bibfnamefont {J.}~\bibnamefont {Dual}},\ }\bibfield
  {title} {\enquote {\bibinfo {title} {Interparticle attraction along the
  direction of the pressure gradient in an acoustic standing wave},}\
  }\href@noop {} {\bibfield  {journal} {\bibinfo  {journal} {Physical Review
  E}\ }\textbf {\bibinfo {volume} {105}},\ \bibinfo {pages} {L053101} (\bibinfo
  {year} {2022})}\BibitemShut {NoStop}%
\bibitem [{\citenamefont {Eslami}\ \emph {et~al.}(2023)\citenamefont {Eslami},
  \citenamefont {Hamzehpour}, \citenamefont {Derikvandi},\ and\ \citenamefont
  {Amir~Bahrani}}]{Eslami2023}%
  \BibitemOpen
  \bibfield  {author} {\bibinfo {author} {\bibfnamefont {F.}~\bibnamefont
  {Eslami}}, \bibinfo {author} {\bibfnamefont {H.}~\bibnamefont {Hamzehpour}},
  \bibinfo {author} {\bibfnamefont {S.}~\bibnamefont {Derikvandi}}, \ and\
  \bibinfo {author} {\bibfnamefont {S.}~\bibnamefont {Amir~Bahrani}},\
  }\bibfield  {title} {\enquote {\bibinfo {title} {Acoustic interaction force
  between two particles immersed in a viscoelastic fluid},}\ }\href {\doibase
  10.1063/5.0143005} {\bibfield  {journal} {\bibinfo  {journal} {Physics of
  Fluids}\ }\textbf {\bibinfo {volume} {35}} (\bibinfo {year} {2023}),\
  10.1063/5.0143005}\BibitemShut {NoStop}%
\bibitem [{\citenamefont {Baasch}, \citenamefont {Leibacher},\ and\
  \citenamefont {Dual}(2017)}]{Baasch2017}%
  \BibitemOpen
  \bibfield  {author} {\bibinfo {author} {\bibfnamefont {T.}~\bibnamefont
  {Baasch}}, \bibinfo {author} {\bibfnamefont {I.}~\bibnamefont {Leibacher}}, \
  and\ \bibinfo {author} {\bibfnamefont {J.}~\bibnamefont {Dual}},\ }\bibfield
  {title} {\enquote {\bibinfo {title} {Multibody dynamics in
  acoustophoresis},}\ }\href {\doibase 10.1121/1.4977030} {\bibfield  {journal}
  {\bibinfo  {journal} {The Journal of the Acoustical Society of America}\
  }\textbf {\bibinfo {volume} {141}},\ \bibinfo {pages} {1664–1674} (\bibinfo
  {year} {2017})}\BibitemShut {NoStop}%
\bibitem [{\citenamefont {St.~Clair}\ \emph {et~al.}(2023)\citenamefont
  {St.~Clair}, \citenamefont {Davenport}, \citenamefont {Kim},\ and\
  \citenamefont {Kleckner}}]{StClair2023}%
  \BibitemOpen
  \bibfield  {author} {\bibinfo {author} {\bibfnamefont {N.}~\bibnamefont
  {St.~Clair}}, \bibinfo {author} {\bibfnamefont {D.}~\bibnamefont
  {Davenport}}, \bibinfo {author} {\bibfnamefont {A.~D.}\ \bibnamefont {Kim}},
  \ and\ \bibinfo {author} {\bibfnamefont {D.}~\bibnamefont {Kleckner}},\
  }\bibfield  {title} {\enquote {\bibinfo {title} {Dynamics of acoustically
  bound particles},}\ }\href {\doibase 10.1103/physrevresearch.5.013051}
  {\bibfield  {journal} {\bibinfo  {journal} {Physical Review Research}\
  }\textbf {\bibinfo {volume} {5}} (\bibinfo {year} {2023}),\
  10.1103/physrevresearch.5.013051}\BibitemShut {NoStop}%
\bibitem [{\citenamefont {Wang}\ \emph {et~al.}(2017)\citenamefont {Wang},
  \citenamefont {Qiu}, \citenamefont {Zhang}, \citenamefont {Han},
  \citenamefont {Ke},\ and\ \citenamefont {Liu}}]{wang2017sound}%
  \BibitemOpen
  \bibfield  {author} {\bibinfo {author} {\bibfnamefont {M.}~\bibnamefont
  {Wang}}, \bibinfo {author} {\bibfnamefont {C.}~\bibnamefont {Qiu}}, \bibinfo
  {author} {\bibfnamefont {S.}~\bibnamefont {Zhang}}, \bibinfo {author}
  {\bibfnamefont {R.}~\bibnamefont {Han}}, \bibinfo {author} {\bibfnamefont
  {M.}~\bibnamefont {Ke}}, \ and\ \bibinfo {author} {\bibfnamefont
  {Z.}~\bibnamefont {Liu}},\ }\bibfield  {title} {\enquote {\bibinfo {title}
  {Sound-mediated stable configurations for polystyrene particles},}\
  }\href@noop {} {\bibfield  {journal} {\bibinfo  {journal} {Physical Review
  E}\ }\textbf {\bibinfo {volume} {96}},\ \bibinfo {pages} {052604} (\bibinfo
  {year} {2017})}\BibitemShut {NoStop}%
\bibitem [{\citenamefont {Silva}\ \emph {et~al.}(2019)\citenamefont {Silva},
  \citenamefont {Lopes}, \citenamefont {Leão-Neto}, \citenamefont {Nichols},\
  and\ \citenamefont {Drinkwater}}]{Silva2019}%
  \BibitemOpen
  \bibfield  {author} {\bibinfo {author} {\bibfnamefont {G.~T.}\ \bibnamefont
  {Silva}}, \bibinfo {author} {\bibfnamefont {J.~H.}\ \bibnamefont {Lopes}},
  \bibinfo {author} {\bibfnamefont {J.~P.}\ \bibnamefont {Leão-Neto}},
  \bibinfo {author} {\bibfnamefont {M.~K.}\ \bibnamefont {Nichols}}, \ and\
  \bibinfo {author} {\bibfnamefont {B.~W.}\ \bibnamefont {Drinkwater}},\
  }\bibfield  {title} {\enquote {\bibinfo {title} {Particle patterning by
  ultrasonic standing waves in a rectangular cavity},}\ }\href {\doibase
  10.1103/physrevapplied.11.054044} {\bibfield  {journal} {\bibinfo  {journal}
  {Physical Review Applied}\ }\textbf {\bibinfo {volume} {11}} (\bibinfo {year}
  {2019}),\ 10.1103/physrevapplied.11.054044}\BibitemShut {NoStop}%
\bibitem [{\citenamefont {Sepehrirahnama}, \citenamefont {Chau},\ and\
  \citenamefont {Lim}(2016)}]{sepehrirahnama2016effects}%
  \BibitemOpen
  \bibfield  {author} {\bibinfo {author} {\bibfnamefont {S.}~\bibnamefont
  {Sepehrirahnama}}, \bibinfo {author} {\bibfnamefont {F.~S.}\ \bibnamefont
  {Chau}}, \ and\ \bibinfo {author} {\bibfnamefont {K.-M.}\ \bibnamefont
  {Lim}},\ }\bibfield  {title} {\enquote {\bibinfo {title} {Effects of
  viscosity and acoustic streaming on the interparticle radiation force between
  rigid spheres in a standing wave},}\ }\href@noop {} {\bibfield  {journal}
  {\bibinfo  {journal} {Physical Review E}\ }\textbf {\bibinfo {volume} {93}},\
  \bibinfo {pages} {023307} (\bibinfo {year} {2016})}\BibitemShut {NoStop}%
\bibitem [{\citenamefont {Collins}\ \emph {et~al.}(2015)\citenamefont
  {Collins}, \citenamefont {Morahan}, \citenamefont {Garcia-Bustos},
  \citenamefont {Doerig}, \citenamefont {Plebanski},\ and\ \citenamefont
  {Neild}}]{collins2015two}%
  \BibitemOpen
  \bibfield  {author} {\bibinfo {author} {\bibfnamefont {D.~J.}\ \bibnamefont
  {Collins}}, \bibinfo {author} {\bibfnamefont {B.}~\bibnamefont {Morahan}},
  \bibinfo {author} {\bibfnamefont {J.}~\bibnamefont {Garcia-Bustos}}, \bibinfo
  {author} {\bibfnamefont {C.}~\bibnamefont {Doerig}}, \bibinfo {author}
  {\bibfnamefont {M.}~\bibnamefont {Plebanski}}, \ and\ \bibinfo {author}
  {\bibfnamefont {A.}~\bibnamefont {Neild}},\ }\bibfield  {title} {\enquote
  {\bibinfo {title} {Two-dimensional single-cell patterning with one cell per
  well driven by surface acoustic waves},}\ }\href@noop {} {\bibfield
  {journal} {\bibinfo  {journal} {Nature communications}\ }\textbf {\bibinfo
  {volume} {6}},\ \bibinfo {pages} {8686} (\bibinfo {year} {2015})}\BibitemShut
  {NoStop}%
\bibitem [{\citenamefont {Nguyen}\ \emph {et~al.}(2018)\citenamefont {Nguyen},
  \citenamefont {Tran}, \citenamefont {Fu},\ and\ \citenamefont
  {Du}}]{nguyen2018patterning}%
  \BibitemOpen
  \bibfield  {author} {\bibinfo {author} {\bibfnamefont {T.~D.}\ \bibnamefont
  {Nguyen}}, \bibinfo {author} {\bibfnamefont {V.~T.}\ \bibnamefont {Tran}},
  \bibinfo {author} {\bibfnamefont {Y.~Q.}\ \bibnamefont {Fu}}, \ and\ \bibinfo
  {author} {\bibfnamefont {H.}~\bibnamefont {Du}},\ }\bibfield  {title}
  {\enquote {\bibinfo {title} {Patterning and manipulating microparticles into
  a three-dimensional matrix using standing surface acoustic waves},}\
  }\href@noop {} {\bibfield  {journal} {\bibinfo  {journal} {Applied Physics
  Letters}\ }\textbf {\bibinfo {volume} {112}} (\bibinfo {year}
  {2018})}\BibitemShut {NoStop}%
\bibitem [{\citenamefont {Mandal}\ and\ \citenamefont
  {Banerjee}(2022)}]{mandal2022surface}%
  \BibitemOpen
  \bibfield  {author} {\bibinfo {author} {\bibfnamefont {D.}~\bibnamefont
  {Mandal}}\ and\ \bibinfo {author} {\bibfnamefont {S.}~\bibnamefont
  {Banerjee}},\ }\bibfield  {title} {\enquote {\bibinfo {title} {Surface
  acoustic wave (saw) sensors: Physics, materials, and applications},}\
  }\href@noop {} {\bibfield  {journal} {\bibinfo  {journal} {Sensors}\ }\textbf
  {\bibinfo {volume} {22}},\ \bibinfo {pages} {820} (\bibinfo {year}
  {2022})}\BibitemShut {NoStop}%
\bibitem [{\citenamefont {Destgeer}\ \emph {et~al.}(2013)\citenamefont
  {Destgeer}, \citenamefont {Lee}, \citenamefont {Jung}, \citenamefont
  {Alazzam},\ and\ \citenamefont {Sung}}]{destgeer2013continuous}%
  \BibitemOpen
  \bibfield  {author} {\bibinfo {author} {\bibfnamefont {G.}~\bibnamefont
  {Destgeer}}, \bibinfo {author} {\bibfnamefont {K.~H.}\ \bibnamefont {Lee}},
  \bibinfo {author} {\bibfnamefont {J.~H.}\ \bibnamefont {Jung}}, \bibinfo
  {author} {\bibfnamefont {A.}~\bibnamefont {Alazzam}}, \ and\ \bibinfo
  {author} {\bibfnamefont {H.~J.}\ \bibnamefont {Sung}},\ }\bibfield  {title}
  {\enquote {\bibinfo {title} {Continuous separation of particles in a pdms
  microfluidic channel via travelling surface acoustic waves (tsaw)},}\
  }\href@noop {} {\bibfield  {journal} {\bibinfo  {journal} {Lab on a Chip}\
  }\textbf {\bibinfo {volume} {13}},\ \bibinfo {pages} {4210--4216} (\bibinfo
  {year} {2013})}\BibitemShut {NoStop}%
\bibitem [{\citenamefont {Toftul}\ \emph {et~al.}(2019)\citenamefont {Toftul},
  \citenamefont {Bliokh}, \citenamefont {Petrov},\ and\ \citenamefont
  {Nori}}]{Toftul2019}%
  \BibitemOpen
  \bibfield  {author} {\bibinfo {author} {\bibfnamefont {I.}~\bibnamefont
  {Toftul}}, \bibinfo {author} {\bibfnamefont {K.}~\bibnamefont {Bliokh}},
  \bibinfo {author} {\bibfnamefont {M.}~\bibnamefont {Petrov}}, \ and\ \bibinfo
  {author} {\bibfnamefont {F.}~\bibnamefont {Nori}},\ }\bibfield  {title}
  {\enquote {\bibinfo {title} {Acoustic radiation force and torque on small
  particles as measures of the canonical momentum and spin densities},}\ }\href
  {\doibase 10.1103/physrevlett.123.183901} {\bibfield  {journal} {\bibinfo
  {journal} {Physical Review Letters}\ }\textbf {\bibinfo {volume} {123}}
  (\bibinfo {year} {2019}),\ 10.1103/physrevlett.123.183901}\BibitemShut
  {NoStop}%
\bibitem [{\citenamefont {Williams}(1999)}]{williams1999fourier}%
  \BibitemOpen
  \bibfield  {author} {\bibinfo {author} {\bibfnamefont {E.~G.}\ \bibnamefont
  {Williams}},\ }\href@noop {} {\emph {\bibinfo {title} {Fourier acoustics:
  sound radiation and nearfield acoustical holography}}}\ (\bibinfo
  {publisher} {Academic press},\ \bibinfo {year} {1999})\BibitemShut {NoStop}%
\bibitem [{\citenamefont {Kostina}\ \emph {et~al.}(2019)\citenamefont
  {Kostina}, \citenamefont {Petrov}, \citenamefont {Ivinskaya}, \citenamefont
  {Sukhov}, \citenamefont {Bogdanov}, \citenamefont {Toftul}, \citenamefont
  {Nieto-Vesperinas}, \citenamefont {Ginzburg},\ and\ \citenamefont
  {Shalin}}]{Kostina2019}%
  \BibitemOpen
  \bibfield  {author} {\bibinfo {author} {\bibfnamefont {N.}~\bibnamefont
  {Kostina}}, \bibinfo {author} {\bibfnamefont {M.}~\bibnamefont {Petrov}},
  \bibinfo {author} {\bibfnamefont {A.}~\bibnamefont {Ivinskaya}}, \bibinfo
  {author} {\bibfnamefont {S.}~\bibnamefont {Sukhov}}, \bibinfo {author}
  {\bibfnamefont {A.}~\bibnamefont {Bogdanov}}, \bibinfo {author}
  {\bibfnamefont {I.}~\bibnamefont {Toftul}}, \bibinfo {author} {\bibfnamefont
  {M.}~\bibnamefont {Nieto-Vesperinas}}, \bibinfo {author} {\bibfnamefont
  {P.}~\bibnamefont {Ginzburg}}, \ and\ \bibinfo {author} {\bibfnamefont
  {A.}~\bibnamefont {Shalin}},\ }\bibfield  {title} {\enquote {\bibinfo {title}
  {Optical binding via surface plasmon polariton interference},}\ }\href
  {\doibase 10.1103/physrevb.99.125416} {\bibfield  {journal} {\bibinfo
  {journal} {Physical Review B}\ }\textbf {\bibinfo {volume} {99}} (\bibinfo
  {year} {2019}),\ 10.1103/physrevb.99.125416}\BibitemShut {NoStop}%
\bibitem [{\citenamefont {Wang}\ and\ \citenamefont {Manmi}(2014)}]{Wang2014}%
  \BibitemOpen
  \bibfield  {author} {\bibinfo {author} {\bibfnamefont {Q.~X.}\ \bibnamefont
  {Wang}}\ and\ \bibinfo {author} {\bibfnamefont {K.}~\bibnamefont {Manmi}},\
  }\bibfield  {title} {\enquote {\bibinfo {title} {Three dimensional
  microbubble dynamics near a wall subject to high intensity ultrasound},}\
  }\href {\doibase 10.1063/1.4866772} {\bibfield  {journal} {\bibinfo
  {journal} {Physics of Fluids}\ }\textbf {\bibinfo {volume} {26}} (\bibinfo
  {year} {2014}),\ 10.1063/1.4866772}\BibitemShut {NoStop}%
\bibitem [{\citenamefont {Baasch}\ and\ \citenamefont
  {Dual}(2020)}]{Baasch2020}%
  \BibitemOpen
  \bibfield  {author} {\bibinfo {author} {\bibfnamefont {T.}~\bibnamefont
  {Baasch}}\ and\ \bibinfo {author} {\bibfnamefont {J.}~\bibnamefont {Dual}},\
  }\bibfield  {title} {\enquote {\bibinfo {title} {Acoustic radiation force on
  a spherical fluid or solid elastic particle placed close to a fluid or solid
  elastic half-space},}\ }\href {\doibase 10.1103/physrevapplied.14.024052}
  {\bibfield  {journal} {\bibinfo  {journal} {Physical Review Applied}\
  }\textbf {\bibinfo {volume} {14}} (\bibinfo {year} {2020}),\
  10.1103/physrevapplied.14.024052}\BibitemShut {NoStop}%
\bibitem [{\citenamefont {K\"{a}hler}, \citenamefont {Platz},\ and\
  \citenamefont {Schmid}(2022)}]{Khler2022}%
  \BibitemOpen
  \bibfield  {author} {\bibinfo {author} {\bibfnamefont {H.}~\bibnamefont
  {K\"{a}hler}}, \bibinfo {author} {\bibfnamefont {D.}~\bibnamefont {Platz}}, \
  and\ \bibinfo {author} {\bibfnamefont {S.}~\bibnamefont {Schmid}},\
  }\bibfield  {title} {\enquote {\bibinfo {title} {Surface acoustic wave
  coupling between micromechanical resonators},}\ }\href {\doibase
  10.1038/s42005-022-00895-2} {\bibfield  {journal} {\bibinfo  {journal}
  {Communications Physics}\ }\textbf {\bibinfo {volume} {5}} (\bibinfo {year}
  {2022}),\ 10.1038/s42005-022-00895-2}\BibitemShut {NoStop}%
\bibitem [{\citenamefont {Bostr\"{o}m}(1980)}]{Bostrm1980}%
  \BibitemOpen
  \bibfield  {author} {\bibinfo {author} {\bibfnamefont {A.}~\bibnamefont
  {Bostr\"{o}m}},\ }\bibfield  {title} {\enquote {\bibinfo {title}
  {Transmission and reflection of acoustic waves by an obstacle in a
  waveguide},}\ }\href {\doibase 10.1016/0165-2125(80)90026-8} {\bibfield
  {journal} {\bibinfo  {journal} {Wave Motion}\ }\textbf {\bibinfo {volume}
  {2}},\ \bibinfo {pages} {167–184} (\bibinfo {year} {1980})}\BibitemShut
  {NoStop}%
\bibitem [{\citenamefont {Gaunaurd}\ and\ \citenamefont
  {Huang}(1994)}]{Gaunaurd1994}%
  \BibitemOpen
  \bibfield  {author} {\bibinfo {author} {\bibfnamefont {G.~C.}\ \bibnamefont
  {Gaunaurd}}\ and\ \bibinfo {author} {\bibfnamefont {H.}~\bibnamefont
  {Huang}},\ }\bibfield  {title} {\enquote {\bibinfo {title} {Acoustic
  scattering by a spherical body near a plane boundary},}\ }\href {\doibase
  10.1121/1.410126} {\bibfield  {journal} {\bibinfo  {journal} {The Journal of
  the Acoustical Society of America}\ }\textbf {\bibinfo {volume} {96}},\
  \bibinfo {pages} {2526–2536} (\bibinfo {year} {1994})}\BibitemShut
  {NoStop}%
\bibitem [{\citenamefont {Shenderov}(2002)}]{Shenderov2002}%
  \BibitemOpen
  \bibfield  {author} {\bibinfo {author} {\bibfnamefont {E.~L.}\ \bibnamefont
  {Shenderov}},\ }\bibfield  {title} {\enquote {\bibinfo {title} {Diffraction
  of sound by an elastic or impedance sphere located near an impedance or
  elastic boundary of a halfspace},}\ }\href {\doibase 10.1134/1.1507206}
  {\bibfield  {journal} {\bibinfo  {journal} {Acoustical Physics}\ }\textbf
  {\bibinfo {volume} {48}},\ \bibinfo {pages} {607–617} (\bibinfo {year}
  {2002})}\BibitemShut {NoStop}%
\bibitem [{\citenamefont {Miri}\ and\ \citenamefont {Mitri}(2011)}]{Miri2011}%
  \BibitemOpen
  \bibfield  {author} {\bibinfo {author} {\bibfnamefont {A.~K.}\ \bibnamefont
  {Miri}}\ and\ \bibinfo {author} {\bibfnamefont {F.~G.}\ \bibnamefont
  {Mitri}},\ }\bibfield  {title} {\enquote {\bibinfo {title} {Acoustic
  radiation force on a spherical contrast agent shell near a vessel porous wall
  – theory},}\ }\href {\doibase 10.1016/j.ultrasmedbio.2010.11.006}
  {\bibfield  {journal} {\bibinfo  {journal} {Ultrasound in Medicine and
  Biology}\ }\textbf {\bibinfo {volume} {37}},\ \bibinfo {pages} {301–311}
  (\bibinfo {year} {2011})}\BibitemShut {NoStop}%
\bibitem [{\citenamefont {Wang}\ and\ \citenamefont {Dual}(2012)}]{Wang2012}%
  \BibitemOpen
  \bibfield  {author} {\bibinfo {author} {\bibfnamefont {J.}~\bibnamefont
  {Wang}}\ and\ \bibinfo {author} {\bibfnamefont {J.}~\bibnamefont {Dual}},\
  }\bibfield  {title} {\enquote {\bibinfo {title} {Theoretical and numerical
  calculation of the acoustic radiation force acting on a circular rigid
  cylinder near a flat wall in a standing wave excitation in an ideal fluid},}\
  }\href {\doibase 10.1016/j.ultras.2011.09.002} {\bibfield  {journal}
  {\bibinfo  {journal} {Ultrasonics}\ }\textbf {\bibinfo {volume} {52}},\
  \bibinfo {pages} {325–332} (\bibinfo {year} {2012})}\BibitemShut {NoStop}%
\bibitem [{\citenamefont {Qiao}, \citenamefont {Zhang},\ and\ \citenamefont
  {Zhang}(2017)}]{Qiao2017}%
  \BibitemOpen
  \bibfield  {author} {\bibinfo {author} {\bibfnamefont {Y.}~\bibnamefont
  {Qiao}}, \bibinfo {author} {\bibfnamefont {X.}~\bibnamefont {Zhang}}, \ and\
  \bibinfo {author} {\bibfnamefont {G.}~\bibnamefont {Zhang}},\ }\bibfield
  {title} {\enquote {\bibinfo {title} {Acoustic radiation force on a fluid
  cylindrical particle immersed in water near an impedance boundary},}\ }\href
  {\doibase 10.1121/1.4986624} {\bibfield  {journal} {\bibinfo  {journal} {The
  Journal of the Acoustical Society of America}\ }\textbf {\bibinfo {volume}
  {141}},\ \bibinfo {pages} {4633–4641} (\bibinfo {year} {2017})}\BibitemShut
  {NoStop}%
\bibitem [{\citenamefont {Mitri}(2018)}]{Mitri2018}%
  \BibitemOpen
  \bibfield  {author} {\bibinfo {author} {\bibfnamefont {F.~G.}\ \bibnamefont
  {Mitri}},\ }\bibfield  {title} {\enquote {\bibinfo {title} {Acoustic
  radiation force on a cylindrical particle near a planar rigid boundary},}\
  }\href {\doibase 10.1088/2399-6528/aab109} {\bibfield  {journal} {\bibinfo
  {journal} {Journal of Physics Communications}\ }\textbf {\bibinfo {volume}
  {2}},\ \bibinfo {pages} {045019} (\bibinfo {year} {2018})}\BibitemShut
  {NoStop}%
\bibitem [{\citenamefont {Zang}\ \emph {et~al.}(2019)\citenamefont {Zang},
  \citenamefont {Qiao}, \citenamefont {Liu},\ and\ \citenamefont
  {Liu}}]{Zang2019}%
  \BibitemOpen
  \bibfield  {author} {\bibinfo {author} {\bibfnamefont {Y.}~\bibnamefont
  {Zang}}, \bibinfo {author} {\bibfnamefont {Y.}~\bibnamefont {Qiao}}, \bibinfo
  {author} {\bibfnamefont {J.}~\bibnamefont {Liu}}, \ and\ \bibinfo {author}
  {\bibfnamefont {X.}~\bibnamefont {Liu}},\ }\bibfield  {title} {\enquote
  {\bibinfo {title} {Axial acoustic radiation force on a fluid sphere between
  two impedance boundaries for gaussian beam},}\ }\href {\doibase
  10.1088/1674-1056/28/3/034301} {\bibfield  {journal} {\bibinfo  {journal}
  {Chinese Physics B}\ }\textbf {\bibinfo {volume} {28}},\ \bibinfo {pages}
  {034301} (\bibinfo {year} {2019})}\BibitemShut {NoStop}%
\bibitem [{\citenamefont {Maksimov}(2020)}]{Maksimov2020}%
  \BibitemOpen
  \bibfield  {author} {\bibinfo {author} {\bibfnamefont {A.}~\bibnamefont
  {Maksimov}},\ }\bibfield  {title} {\enquote {\bibinfo {title} {Splitting of
  the surface modes for bubble oscillations near a boundary},}\ }\href
  {\doibase 10.1063/5.0025196} {\bibfield  {journal} {\bibinfo  {journal}
  {Physics of Fluids}\ }\textbf {\bibinfo {volume} {32}} (\bibinfo {year}
  {2020}),\ 10.1063/5.0025196}\BibitemShut {NoStop}%
\bibitem [{\citenamefont {Simon}\ and\ \citenamefont
  {Hamilton}(2023{\natexlab{a}})}]{Simon2023}%
  \BibitemOpen
  \bibfield  {author} {\bibinfo {author} {\bibfnamefont {B.~E.}\ \bibnamefont
  {Simon}}\ and\ \bibinfo {author} {\bibfnamefont {M.~F.}\ \bibnamefont
  {Hamilton}},\ }\bibfield  {title} {\enquote {\bibinfo {title} {Analytical
  solution for acoustic radiation force on a sphere near a planar boundary},}\
  }\href {\doibase 10.1121/10.0016885} {\bibfield  {journal} {\bibinfo
  {journal} {The Journal of the Acoustical Society of America}\ }\textbf
  {\bibinfo {volume} {153}},\ \bibinfo {pages} {627–642} (\bibinfo {year}
  {2023}{\natexlab{a}})}\BibitemShut {NoStop}%
\bibitem [{\citenamefont {Doinikov}\ and\ \citenamefont
  {Bouakaz}(2015)}]{Doinikov2015}%
  \BibitemOpen
  \bibfield  {author} {\bibinfo {author} {\bibfnamefont {A.~A.}\ \bibnamefont
  {Doinikov}}\ and\ \bibinfo {author} {\bibfnamefont {A.}~\bibnamefont
  {Bouakaz}},\ }\bibfield  {title} {\enquote {\bibinfo {title} {Interaction of
  an ultrasound-activated contrast microbubble with a wall at arbitrary
  separation distances},}\ }\href {\doibase 10.1088/0031-9155/60/20/7909}
  {\bibfield  {journal} {\bibinfo  {journal} {Physics in Medicine and Biology}\
  }\textbf {\bibinfo {volume} {60}},\ \bibinfo {pages} {7909–7925} (\bibinfo
  {year} {2015})}\BibitemShut {NoStop}%
\bibitem [{\citenamefont {Westervelt}(1957)}]{Westervelt1957}%
  \BibitemOpen
  \bibfield  {author} {\bibinfo {author} {\bibfnamefont {P.~J.}\ \bibnamefont
  {Westervelt}},\ }\bibfield  {title} {\enquote {\bibinfo {title} {Acoustic
  radiation pressure},}\ }\href {\doibase 10.1121/1.1908669} {\bibfield
  {journal} {\bibinfo  {journal} {The Journal of the Acoustical Society of
  America}\ }\textbf {\bibinfo {volume} {29}},\ \bibinfo {pages} {26–29}
  (\bibinfo {year} {1957})}\BibitemShut {NoStop}%
\bibitem [{\citenamefont {Sapozhnikov}\ and\ \citenamefont
  {Bailey}(2013)}]{sapozhnikov2013radiation}%
  \BibitemOpen
  \bibfield  {author} {\bibinfo {author} {\bibfnamefont {O.~A.}\ \bibnamefont
  {Sapozhnikov}}\ and\ \bibinfo {author} {\bibfnamefont {M.~R.}\ \bibnamefont
  {Bailey}},\ }\bibfield  {title} {\enquote {\bibinfo {title} {Radiation force
  of an arbitrary acoustic beam on an elastic sphere in a fluid},}\ }\href@noop
  {} {\bibfield  {journal} {\bibinfo  {journal} {The Journal of the Acoustical
  Society of America}\ }\textbf {\bibinfo {volume} {133}},\ \bibinfo {pages}
  {661--676} (\bibinfo {year} {2013})}\BibitemShut {NoStop}%
\bibitem [{\citenamefont {de~Vries}, \citenamefont {van Coevorden},\ and\
  \citenamefont {Lagendijk}(1998)}]{DeVries1998}%
  \BibitemOpen
  \bibfield  {author} {\bibinfo {author} {\bibfnamefont {P.}~\bibnamefont
  {de~Vries}}, \bibinfo {author} {\bibfnamefont {D.}~\bibnamefont {van
  Coevorden}}, \ and\ \bibinfo {author} {\bibfnamefont {A.}~\bibnamefont
  {Lagendijk}},\ }\bibfield  {title} {\enquote {\bibinfo {title} {Point
  scatterers for classical waves},}\ }\href {\doibase
  10.1103/RevModPhys.70.447} {\bibfield  {journal} {\bibinfo  {journal}
  {Reviews of Modern Physics}\ }\textbf {\bibinfo {volume} {70}},\ \bibinfo
  {pages} {447--466} (\bibinfo {year} {1998})},\ \bibinfo {note} {iSBN:
  0034-6861}\BibitemShut {NoStop}%
\bibitem [{\citenamefont {Maurice~Ewing}(1957)}]{maurice_ewing_elastic_1957}%
  \BibitemOpen
  \bibfield  {author} {\bibinfo {author} {\bibfnamefont {W.}~\bibnamefont
  {Maurice~Ewing}},\ }\href
  {http://archive.org/details/elasticwavesinla032682mbp} {{\emph {\bibinfo {title} {Elastic {Waves} {In} {Layered} {Media}}}}}\
  (\bibinfo  {publisher} {McGraw Hill Book Company Inc.},\ \bibinfo {year}
  {1957})\BibitemShut {NoStop}%
\bibitem [{\citenamefont {Royer}\ and\ \citenamefont
  {Dieulesaint}(1999)}]{royer1999elastic}%
  \BibitemOpen
  \bibfield  {author} {\bibinfo {author} {\bibfnamefont {D.}~\bibnamefont
  {Royer}}\ and\ \bibinfo {author} {\bibfnamefont {E.}~\bibnamefont
  {Dieulesaint}},\ }\href@noop {} {\emph {\bibinfo {title} {Elastic waves in
  solids I: Free and guided propagation}}}\ (\bibinfo  {publisher} {Springer
  Science \& Business Media},\ \bibinfo {year} {1999})\BibitemShut {NoStop}%
\bibitem [{\citenamefont {Bertoni}\ and\ \citenamefont
  {Tamir}(1973)}]{bertoni1973unified}%
  \BibitemOpen
  \bibfield  {author} {\bibinfo {author} {\bibfnamefont {H.}~\bibnamefont
  {Bertoni}}\ and\ \bibinfo {author} {\bibfnamefont {T.}~\bibnamefont
  {Tamir}},\ }\bibfield  {title} {\enquote {\bibinfo {title} {Unified theory of
  rayleigh-angle phenomena for acoustic beams at liquid-solid interfaces},}\
  }\href@noop {} {\bibfield  {journal} {\bibinfo  {journal} {Applied physics}\
  }\textbf {\bibinfo {volume} {2}},\ \bibinfo {pages} {157--172} (\bibinfo
  {year} {1973})}\BibitemShut {NoStop}%
\bibitem [{\citenamefont {Gedge}\ and\ \citenamefont
  {Hill}(2012)}]{gedge2012acoustofluidics}%
  \BibitemOpen
  \bibfield  {author} {\bibinfo {author} {\bibfnamefont {M.}~\bibnamefont
  {Gedge}}\ and\ \bibinfo {author} {\bibfnamefont {M.}~\bibnamefont {Hill}},\
  }\bibfield  {title} {\enquote {\bibinfo {title} {Acoustofluidics 17: Theory
  and applications of surface acoustic wave devices for particle
  manipulation},}\ }\href@noop {} {\bibfield  {journal} {\bibinfo  {journal}
  {Lab on a Chip}\ }\textbf {\bibinfo {volume} {12}},\ \bibinfo {pages}
  {2998--3007} (\bibinfo {year} {2012})}\BibitemShut {NoStop}%
\bibitem [{\citenamefont {Garcia-Sabat{\'e}}\ \emph {et~al.}(2014)\citenamefont
  {Garcia-Sabat{\'e}}, \citenamefont {Castro}, \citenamefont {Hoyos},\ and\
  \citenamefont {Gonz{\'a}lez-Cinca}}]{garcia2014experimental}%
  \BibitemOpen
  \bibfield  {author} {\bibinfo {author} {\bibfnamefont {A.}~\bibnamefont
  {Garcia-Sabat{\'e}}}, \bibinfo {author} {\bibfnamefont {A.}~\bibnamefont
  {Castro}}, \bibinfo {author} {\bibfnamefont {M.}~\bibnamefont {Hoyos}}, \
  and\ \bibinfo {author} {\bibfnamefont {R.}~\bibnamefont
  {Gonz{\'a}lez-Cinca}},\ }\bibfield  {title} {\enquote {\bibinfo {title}
  {Experimental study on inter-particle acoustic forces},}\ }\href@noop {}
  {\bibfield  {journal} {\bibinfo  {journal} {The Journal of the Acoustical
  Society of America}\ }\textbf {\bibinfo {volume} {135}},\ \bibinfo {pages}
  {1056--1063} (\bibinfo {year} {2014})}\BibitemShut {NoStop}%
\bibitem [{\citenamefont {Saeidi}\ \emph {et~al.}(2019)\citenamefont {Saeidi},
  \citenamefont {Saghafian}, \citenamefont {Haghjooy~Javanmard}, \citenamefont
  {Hammarstr{\"o}m},\ and\ \citenamefont {Wiklund}}]{saeidi2019acoustic}%
  \BibitemOpen
  \bibfield  {author} {\bibinfo {author} {\bibfnamefont {D.}~\bibnamefont
  {Saeidi}}, \bibinfo {author} {\bibfnamefont {M.}~\bibnamefont {Saghafian}},
  \bibinfo {author} {\bibfnamefont {S.}~\bibnamefont {Haghjooy~Javanmard}},
  \bibinfo {author} {\bibfnamefont {B.}~\bibnamefont {Hammarstr{\"o}m}}, \ and\
  \bibinfo {author} {\bibfnamefont {M.}~\bibnamefont {Wiklund}},\ }\bibfield
  {title} {\enquote {\bibinfo {title} {Acoustic dipole and monopole effects in
  solid particle interaction dynamics during acoustophoresis},}\ }\href@noop {}
  {\bibfield  {journal} {\bibinfo  {journal} {The Journal of the Acoustical
  Society of America}\ }\textbf {\bibinfo {volume} {145}},\ \bibinfo {pages}
  {3311--3319} (\bibinfo {year} {2019})}\BibitemShut {NoStop}%
\bibitem [{\citenamefont {Simon}\ and\ \citenamefont
  {Hamilton}(2023{\natexlab{b}})}]{simon2023measurement}%
  \BibitemOpen
  \bibfield  {author} {\bibinfo {author} {\bibfnamefont {B.~E.}\ \bibnamefont
  {Simon}}\ and\ \bibinfo {author} {\bibfnamefont {M.~F.}\ \bibnamefont
  {Hamilton}},\ }\bibfield  {title} {\enquote {\bibinfo {title} {Measurement of
  acoustic radiation force on a sphere near a boundary},}\ }in\ \href@noop {}
  {\emph {\bibinfo {booktitle} {Proceedings of Meetings on Acoustics}}},\
  Vol.~\bibinfo {volume} {51}\ (\bibinfo {organization} {AIP Publishing},\
  \bibinfo {year} {2023})\BibitemShut {NoStop}%
\bibitem [{\citenamefont {Nikolaeva}, \citenamefont {Tsysar},\ and\
  \citenamefont {Sapozhnikov}(2016)}]{nikolaeva2016measuring}%
  \BibitemOpen
  \bibfield  {author} {\bibinfo {author} {\bibfnamefont {A.}~\bibnamefont
  {Nikolaeva}}, \bibinfo {author} {\bibfnamefont {S.}~\bibnamefont {Tsysar}}, \
  and\ \bibinfo {author} {\bibfnamefont {O.}~\bibnamefont {Sapozhnikov}},\
  }\bibfield  {title} {\enquote {\bibinfo {title} {Measuring the radiation
  force of megahertz ultrasound acting on a solid spherical scatterer},}\
  }\href@noop {} {\bibfield  {journal} {\bibinfo  {journal} {Acoustical
  Physics}\ }\textbf {\bibinfo {volume} {62}},\ \bibinfo {pages} {38--45}
  (\bibinfo {year} {2016})}\BibitemShut {NoStop}%
\bibitem [{\citenamefont {Bliokh}\ and\ \citenamefont
  {Nori}(2019)}]{bliokh2019transverse}%
  \BibitemOpen
  \bibfield  {author} {\bibinfo {author} {\bibfnamefont {K.~Y.}\ \bibnamefont
  {Bliokh}}\ and\ \bibinfo {author} {\bibfnamefont {F.}~\bibnamefont {Nori}},\
  }\bibfield  {title} {\enquote {\bibinfo {title} {Transverse spin and surface
  waves in acoustic metamaterials},}\ }\href@noop {} {\bibfield  {journal}
  {\bibinfo  {journal} {Physical Review B}\ }\textbf {\bibinfo {volume} {99}},\
  \bibinfo {pages} {020301} (\bibinfo {year} {2019})}\BibitemShut {NoStop}%
\bibitem [{\citenamefont {Park}\ \emph {et~al.}(2011)\citenamefont {Park},
  \citenamefont {Park}, \citenamefont {Lee}, \citenamefont {Seo}, \citenamefont
  {Kim},\ and\ \citenamefont {Lee}}]{park2011amplification}%
  \BibitemOpen
  \bibfield  {author} {\bibinfo {author} {\bibfnamefont {C.~M.}\ \bibnamefont
  {Park}}, \bibinfo {author} {\bibfnamefont {J.~J.}\ \bibnamefont {Park}},
  \bibinfo {author} {\bibfnamefont {S.~H.}\ \bibnamefont {Lee}}, \bibinfo
  {author} {\bibfnamefont {Y.~M.}\ \bibnamefont {Seo}}, \bibinfo {author}
  {\bibfnamefont {C.~K.}\ \bibnamefont {Kim}}, \ and\ \bibinfo {author}
  {\bibfnamefont {S.~H.}\ \bibnamefont {Lee}},\ }\bibfield  {title} {\enquote
  {\bibinfo {title} {Amplification of acoustic evanescent waves using
  metamaterial slabs},}\ }\href@noop {} {\bibfield  {journal} {\bibinfo
  {journal} {Physical review letters}\ }\textbf {\bibinfo {volume} {107}},\
  \bibinfo {pages} {194301} (\bibinfo {year} {2011})}\BibitemShut {NoStop}%
\bibitem [{\citenamefont {Ambati}\ \emph {et~al.}(2007)\citenamefont {Ambati},
  \citenamefont {Fang}, \citenamefont {Sun},\ and\ \citenamefont
  {Zhang}}]{ambati2007surface}%
  \BibitemOpen
  \bibfield  {author} {\bibinfo {author} {\bibfnamefont {M.}~\bibnamefont
  {Ambati}}, \bibinfo {author} {\bibfnamefont {N.}~\bibnamefont {Fang}},
  \bibinfo {author} {\bibfnamefont {C.}~\bibnamefont {Sun}}, \ and\ \bibinfo
  {author} {\bibfnamefont {X.}~\bibnamefont {Zhang}},\ }\bibfield  {title}
  {\enquote {\bibinfo {title} {Surface resonant states and superlensing in
  acoustic metamaterials},}\ }\href@noop {} {\bibfield  {journal} {\bibinfo
  {journal} {Physical Review B—Condensed Matter and Materials Physics}\
  }\textbf {\bibinfo {volume} {75}},\ \bibinfo {pages} {195447} (\bibinfo
  {year} {2007})}\BibitemShut {NoStop}%
\end{thebibliography}%

\end{document}